# Spectrally-Shaped Continuous-Variable QKD Operating at 500 MHz Over an Optical Pipe Lit by 11 DWDM Channels

**Dinka Milovančev, Nemanja Vokić, Fabian Laudenbach,**
**Christoph Pacher, Hannes Hübel, and Bernhard Schrenk**
*AIT Austrian Institute of Technology, Center for Digital Safety&Security / Security & Communication Technologies, 1210 Vienna, Austria.*
*Author e-mail address: bernhard.schrenk@ait.ac.at*

We demonstrate high-rate CV-QKD supporting a secure-key rate of 22Mb/s through spectral tailoring and optimal use of quantum receiver bandwidth. Co-existence with 11 adjacent carrier-grade C-band channels spaced by only 20nm is accomplished at >10Mb/s.

## 1. Introduction

Quantum key distribution (QKD) promises immunity to eavesdropping in the age of quantum computing, which would render classic public-key cryptography defenseless. In contrast to commonly used discrete-variable QKD, continuous-variable (CV) QKD has been identified as an attractive alternative. It encodes the quantum states into weak coherent states using modulation in quadrature space [1] and omits the need for special purpose single-photon receivers since standard opto-electronics, as commonly found in the existing optical fiber systems, can be applied instead. Furthermore, electro-optical filtering inherent to the coherent detection mechanism reduces the sensitivity to Raman noise of potential co-existing classical channels [2,3]. However, practical deployment dictates that the local oscillator (LO) for the coherent receiver is to be generated at the receiver side, and thus needs to be carefully synchronized. Although this task is well known from classical telecommunication systems, it becomes challenging when operating at very low signal power levels as required for quantum communication.

In this work we demonstrate a spectrally shaped C-band CV-QKD system that operates at a 500-MHz symbol rate. Free-running LO operation is accomplished through a polarization- and frequency-multiplexed pilot-tone scheme with carrier- and sideband-suppressed modulation. We show that secure-key rates of 22.3 and 12 Mb/s are feasible over a 13.2 km link reach without and with co-propagation of classical channels in the same waveband.

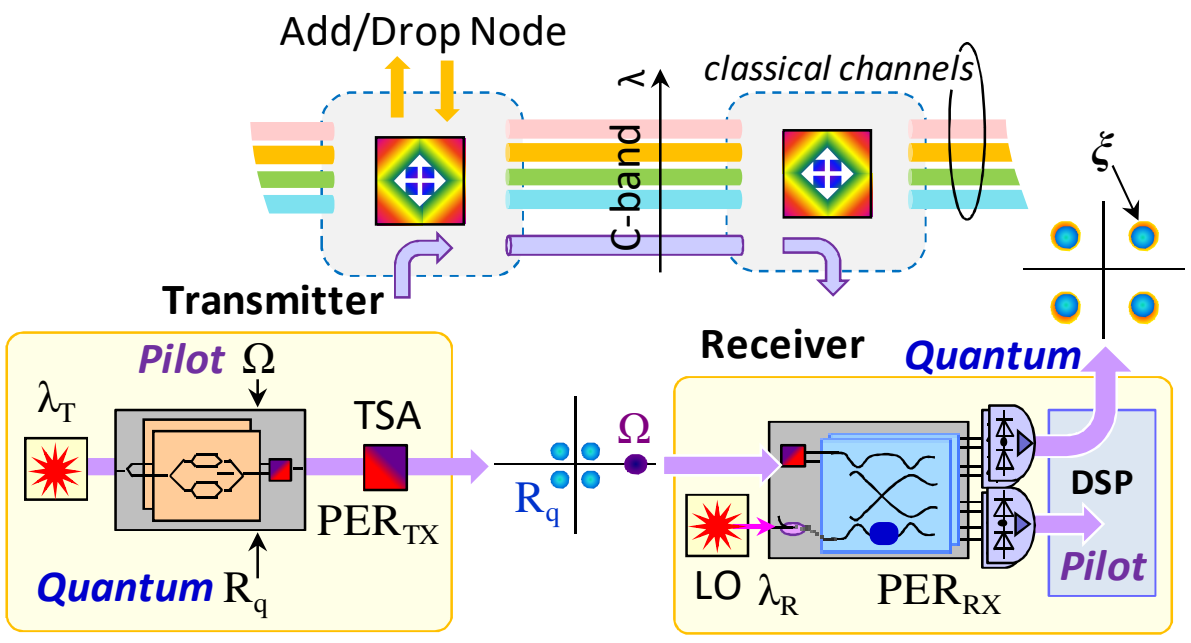

Fig. 1. CV-QKD integrated in lit C-band fiber link.

## 2. CV-QKD System with Polarization-Multiplexed Composite Signal

The robustness of CV-QKD links supports co-existence with carrier-grade DWDM channels in point-to-point links, as sketched in Fig. 1. CV links can be added and dropped between two network nodes, re-using the same spectral waveband as the classical channels. To synchronize the sources at CV transmitter ($\lambda_T$) and receiver ($\lambda_R$) in terms of optical frequency and phase, a pilot tone at frequency Ω is multiplexed in polarization to the quantum signal, which is sent in a discrete 4-state protocol at symbol rate $R_q$ < Ω, thus the pilot also falls out-of-band. As demonstrated earlier [4], optically carrier-suppressed (oCS) single-sideband (SSB) modulation ensures minimal crosstalk of the strong pilot tone to the quantum signal, which can then be recovered in DSP-assisted coherent intradyne detection.

Figure 2 presents the experimental setup. The source at $\lambda_T$ = 1532.9 nm feeds a $LiNbO_3$ I/Q modulator where the pilot and quantum data are encoded and polarization division multiplexed (PDM). The four quantum states are transmitted as QPSK signal in the TE polarization at $R_q$ = 250 MHz for Gaussian and $R_q$ = 500 MHz for Nyquist pulse shaping. The TM polarization carries the pilot tone, which trains the receiver. Moreover, optical pulse carving (CRV) is precedingly applied to the compound signal. The desired power difference between quantum signal and pilot tone is achieved through tributary-selective attenuation (TSA). It is accomplished through leveling the quantum polarization tributary ($PBS_1$, $A_q$, $PBS_2$), by the polarization extinction ratio ($PER_{TX}$) of 20 dB. The spectral displacement of the pilot tone at $\Omega$ = 1 GHz ensures that a relatively strong launch of -46 dBm can be supported without detrimental crosstalk to the quantum channel, provided that the polarization tributaries can be demultiplexed with high extinction ($PER_{RX}$) at the receiver. The eye diagram at the quantum plane can be observed at the intensity monitor *qMon*. Figure 3a presents the eyes at the quantum tributary for uncarved (α) and carved (b) Gaussian-shaped QPSK and uncarved Nyquist-shaped QPSK. The heterodyned RF spectrum at the pilot plane is acquired through monitor *πMon* (δ in Fig. 3b). Polarization alignment at $PC_1/PBS_1$ and thus high $PER_{TX}$ is accomplished by minimizing quantum data at the heterodyne frequency IF in the pilot plane. oCS-SSB modulation is performed to minimize pilot crosstalk at the quantum receiver branch, thus ensuring a low excess noise ξ. Figure 3b indicates a high oCS of 23 dB and an optical sideband suppression of 14 dB at the pilot mirror frequency -Ω.

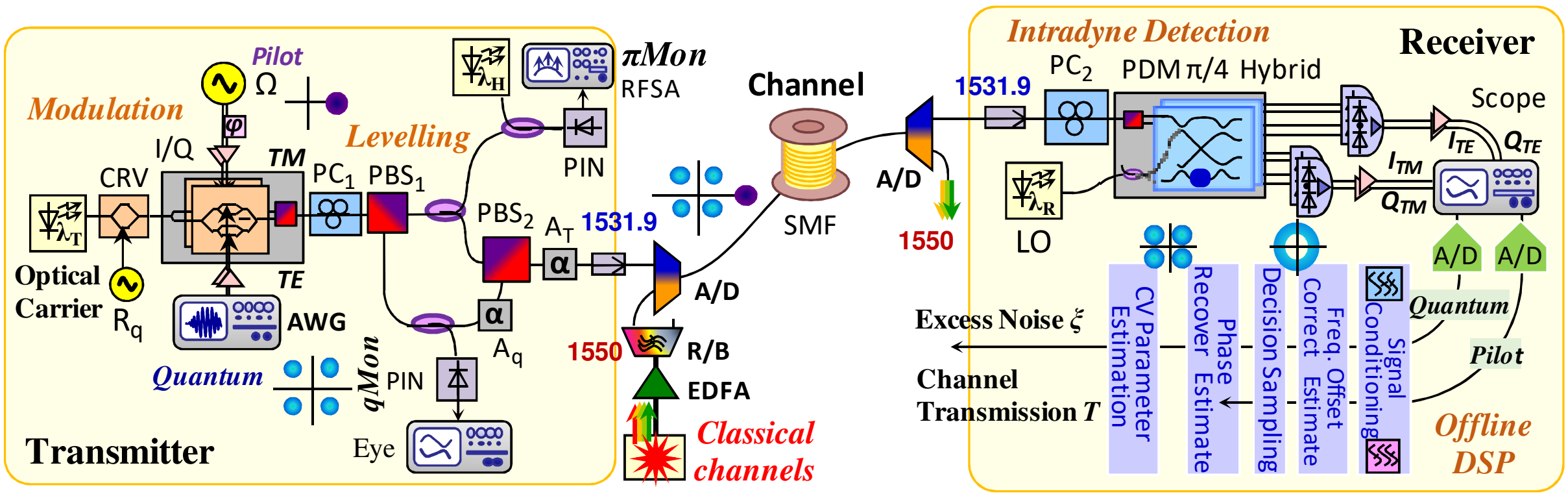

Fig. 2. Experimental setup for CV-QKD system over loaded point-to-point link where classical signals co-exist with the quantum channel.

The compound signal after polarization multiplexing at $PBS_2$ is launched with a desired quantum launch power of 4 photons/symbol and transmitted over a standard SMF channel with a reach of 13.2 and 28.4 km. A C-band comb comprising of 11 classical channels (Fig. 3c) with a launch of 5.2 dBm/λ was co-propagated with the QKD channel. For this purpose the classical channels have been cleaned in its far-reaching spectral tails by a red/blue (R/B) filter and multiplexed to the quantum channel by an add/drop (A/D) filter centered at 1531.9 nm.

At the receiving site an A/D filter drops the quantum signal to a PDM coherent intradyne receiver with free-running LO. Low-linewidth (10 kHz) lasers were employed at transmitter and receiver, with an absolute wavelength stability that ensures that the maximum LO offset from $\lambda_T$ remained within 10 MHz. The delivered polarization was adjusted manually ($PC_2$) before the PDM π/4 hybrid. Balanced PIN photoreceiver pairs with a bandwidth of 315 MHz and 10 GHz were used at the quantum and pilot plane, respectively. Frequency offset estimation/correction, carrier-phase acquisition and recovery (CPR) were performed off-line after digitization of the received signals with a real-time oscilloscope. The DSP stack exploits the acquired frequency and phase information of the strong pilot tone with high signal-to-noise ratio to apply respective corrections subsequently to the quantum data. The parameter estimation of the demodulated quantum data finally yields the excess noise ξ and the secure-key rate $R_S$.

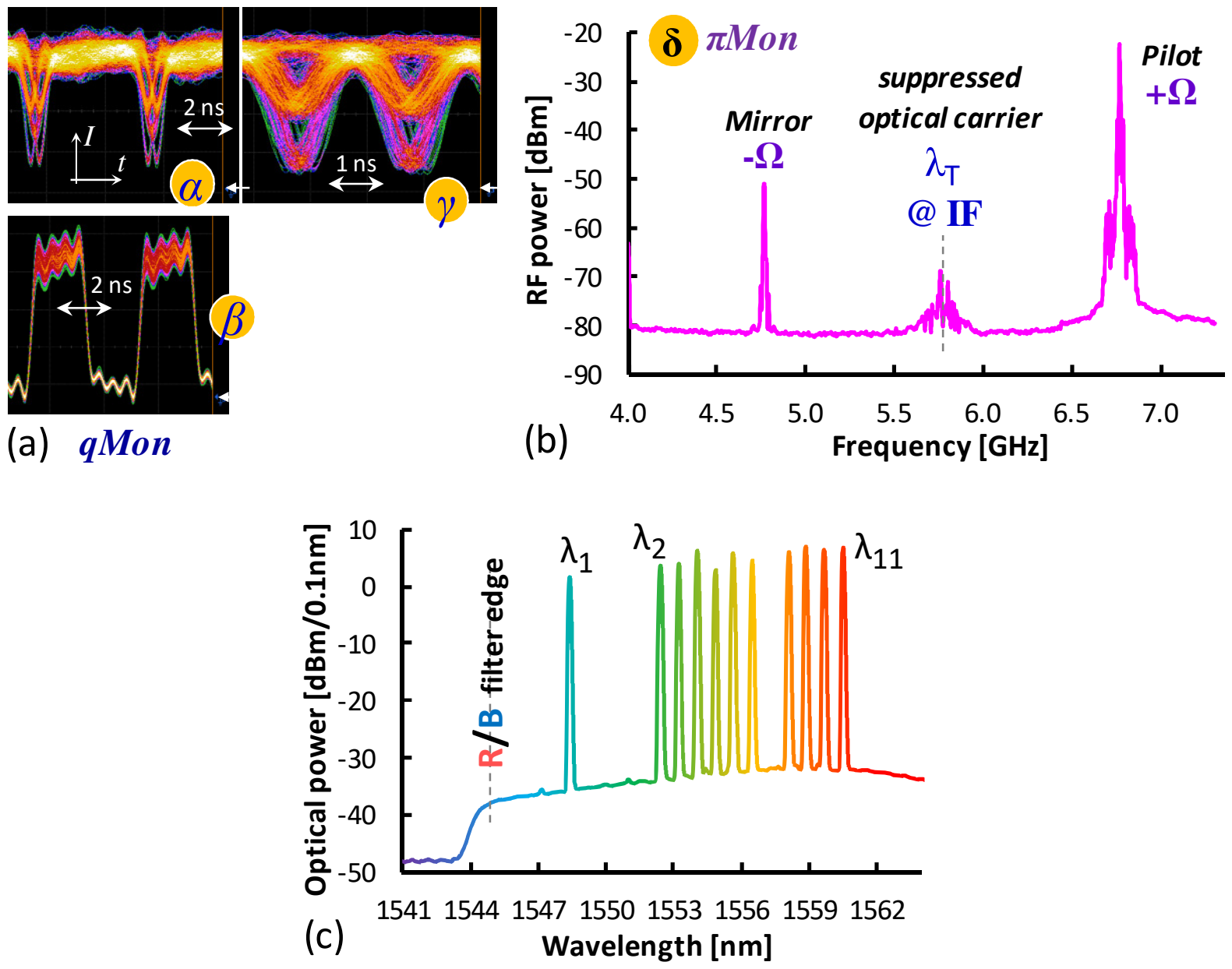

Fig. 3. Transmitter-side monitor signals for (a) quantum signal and (b) pilot tone.
(c) Optical spectrum of classical C-band network load.

## 3. CV-QKD Performance: Spectral Tailoring and Co-Existence with Classical Channels

The received signal spectra in the quantum and pilot plane are presented in Fig. 4a for optimal polarization alignment. The crosstalk of the pilot tone to the quantum plane ($\Psi$) is small, yet it can serve the purpose of polarization tracking. The spectral tailoring through the digital Nyquist pulse shaping filter $H_{RC}$ can be seen in the received quantum signal, leading to an optimal use of available reception bandwidth. The crosstalk note at the mirror pilot frequency is kept small in the pilot plane, meaning an accurate training for the CV-QKD receiver. The estimated pilot phase showed a maximum phase change of 1.75 rad/μs.

Using the methods described in [5], a parameter extraction was performed on the recovered quantum data. This yields the excess noise of the quantum data. The total excess noise $\xi$ contains the quantum shot-noise due to LO and the quadrature variance. The case when the receiver is considered as being hosted at a safe location is also considered. The latter case gives a relaxed security constraint since the shot-noise and the noise of the receiver electrical circuitry are omitted from the total noise, resulting in $\xi_S$ [5].

Table 1 summarizes the results for optimized DSP settings. Results are provided for a transmission reach of 13 km in terms of excess noise $\xi$ and $\xi_S$, which are expressed in quantum shot-noise units (SNU), and the signal-to-noise ratio (SNR). In the case of Nyquist pulse shaping at twice the symbol rate, slightly higher $\xi$ and $\xi_S$ values were obtained, as it is expected from the extended bandwidth.

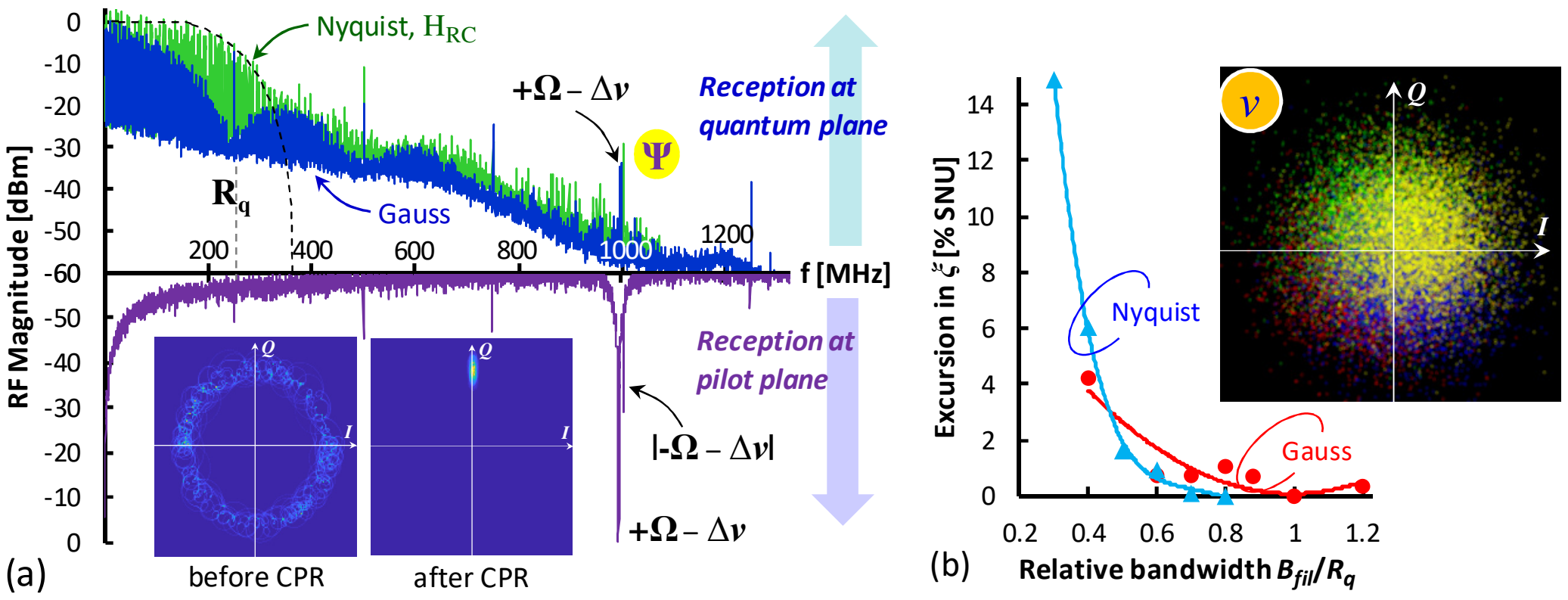

Fig. 4. (a) Received signal spectra in quantum and pilot plane. (b) Dependence of excess noise excursion on relative bandwidth.

DSP parameters such as decision sampling accuracy and filter bandwidth ($B_{fil}$) show a strong influence on the results. Figure 4b reports this dependence of $B_{fil}/R_q$ on $\xi$ for both, Gaussian ($R_q$ = 250 MHz) and Nyquist pulse shaping ($R_q$ = 500 MHz). For Gaussian shaping, an excess reception bandwidth in the digital domain leads to an increasing $\xi$ penalty. In case of Nyquist shaping, this effect is less pronounced since an increase of $B_{fil}$ beyond the 3-dB quantum receiver bandwidth does not equivalently accumulate noise. Yet, Nyquist pulse shaping was more sensitive on the sampling accuracy, which is evident from the reduced eye width (Fig. 3a, $\gamma$). A recovered Nyquist-shaped QPSK constellation is shown as inset in Fig. 4b ($\nu$).

The supported secure-key rates $R_S$ were estimated under the conditions of a Gaussian modulation alphabet and a reconciliation efficiency ($\beta = 0.95$). If the total excess noise $\xi$ is considered, no secure key can be obtained. However, in practical deployments the receiver location is usually designated as a secure one and $\xi_S$ can be adopted. This realistic assumption gives $R_S$ = 22.3 and 12 Mbit/s for Nyquist pulse-shaped CV-QKD without and with co-existence with 11 classical DWDM channels in the same transmission waveband, respectively. Gaussian pulse shaping at the reduced symbol rate yields lower key rates in the range of 10 Mbit/s. At a longer transmission span of 28.4 km, a reduced key rate of 1.43 Mbit/s is obtained without co-propagation of classical channels.

| Pulse / Property | Gaussian-shaped | | | Nyquist-shaped | |
|---|---|---|---|---|---|
| Pulse carving | none | applied | | | |
| **classical ch.** | **0** | **0** | **11λ** | **0** | **11λ** |
| SNR [1] | 0.38 | 0.28 | 0.12 | 0.16 | 0.17 |
| $\xi$ [%SNU] | 1.465 | 1.446 | 1.683 | 1.721 | 2.12 |
| $\xi_S$ [%SNU] | **0.111** | **0.092** | **0.329** | **0.115** | **0.514** |
| $R_S$ [Mb/s] | 11.2 | 11.5 | 8.1 | 22.3 | 12 |

Table 1. CV-QKD performance for a link reach of 13.2 km.

## 4. Conclusions

We have demonstrated high-rate QKD link integration with LO synchronization at the receiver site through a pilot-tone assisted CV system. A high symbol rate of 500 MHz is supported through spectral tailoring of the quantum signal, despite the bandwidth limitations of state-of-the-art low-noise quantum receivers. Secure-key rates of 22 Mb/s or $4.4 \times 10^{-2}$ bits/pulse have been estimated for a link reach of 13.2 km. Co-existence of the C-band QKD link with carrier-grade DWDM channels spaced by ~20 nm has been accomplished, rendering secure-key rates of >10 Mb/s as feasible in this a challenging deployment scenario.

## 5. Acknowledgement

This work has received funding from the EU Horizon-2020 research and innovation programme under grant agreement No 820466 and was further supported by the European Research Council under grant agreement No 804769.